\documentstyle[epsfig]{aipproc}
\def\HI{\hbox{H~$\scriptstyle\rm I\ $}}
\def\HII{\hbox{H~$\scriptstyle\rm II\ $}}

\def\nH{{\rm H}}
\def\nHII{{\rm HII}}

\def\kms{\,{\rm km\,s^{-1}}}
\def\kmsmpc{\,{\rm km\,s^{-1}\,Mpc^{-1}}}

\def\uvunits{\,{\rm ergs\,cm^{-2}\,s^{-1}\,Hz^{-1}\,sr^{-1}}}
\def\eblunits{\,{\rm nW\,m^{-2}\,sr^{-1}}}

\def\ndotunits{\,{\rm s^{-1}\,Mpc^{-3}}}
\def\ndotun{\,{\rm phot\,s^{-1}}}
\def\msun{\,{\rm M_\odot}}
\def\mden{\,{\rm M_\odot\,Mpc^{-3}}}
\def\sfrd{\,{\rm M_\odot\,yr^{-1}\,Mpc^{-3}}}
\def\sfr{\,{\rm M_\odot\,yr^{-1}}}

\def\Lya{Ly$\alpha\ $}
\def\etal{{et al.\ }}

\def\spose#1{\hbox to 0pt{#1\hss}}
\def\lta{\mathrel{\spose{\lower 3pt\hbox{$\mathchar"218$}}
     \raise 2.0pt\hbox{$\mathchar"13C$}}}
\def\gta{\mathrel{\spose{\lower 3pt\hbox{$\mathchar"218$}}
     \raise 2.0pt\hbox{$\mathchar"13E$}}}

\begin{document}
\title{Starlight in the Universe \footnote{To appear in Physica Scripta, 
Proceedings of the Nobel Symposium, Particle Physics and the 
Universe (Enkoping, Sweden, August 20-25, 1998).}}

\author{Piero Madau}
\address{\it Space Telescope Science Institute, Baltimore, MD 21218}
\maketitle

\begin{abstract}
There has been remarkable progress recently in both observational and 
theoretical studies of galaxy formation and evolution. Largely 
due to a combination of deep {\it Hubble Space Telescope} ({\it HST}) imaging, 
Keck spectroscopy, and {\it COBE} far-IR background measurements, 
new constraints have emerged on the emission history of the galaxy population 
as a whole. The global ultraviolet, optical, 
near- and far-IR photometric properties of the universe as a function of 
cosmic time cannot be reproduced by a simple stellar evolution model defined 
by a constant (comoving) star-formation density and a universal (Salpeter) 
initial mass function, and require instead a substantial increase in the 
stellar birthrate with lookback time. While the bulk of the stars present 
today appears to have formed relatively recently, the existence of a decline 
in the star-formation density above $z\approx 2$ remains uncertain. The 
study of the transition from the cosmic `dark age' to an ionized universe 
populated with luminous sources can shed new light on the star formation 
activity at high redshifts, and promises answers to some fundamental 
questions on the formation of cosmic structures.   
If stellar sources are responsible for photoionizing the 
intergalactic medium at $z\approx 5$, the rate of star formation at this 
epoch must be comparable or greater than the one inferred from optical 
observations of galaxies at $z\approx 3$. A population of quasars  
at $z\lta 2$ could make a significant contribution to the extragalactic 
background light if dust-obscured accretion onto supermassive black 
holes is an efficient process.

\end{abstract}

\section*{Introduction}

There is little doubt that the last few years have been very exciting times in 
galaxy formation and evolution studies. The remarkable progress in our 
understanding of faint galaxy data made possible by the combination of 
{\it Hubble Space Telescope} {(\it HST}) deep imaging and 
ground-based spectroscopy has permitted to shed new light on the evolution of 
the stellar birthrate in the
universe, to identify the epoch $1\lta z\lta 2$ where most of 
the optical extragalactic background light was produced, and to set important
contraints on galaxy evolution scenarios. The 
explosion in the quantity of information available on the high-redshift 
universe at optical wavelengths has been complemented by the detection of 
the far-IR/sub-mm background by DIRBE and FIRAS onboard the {\it COBE}
satellite, and by theoretical progress
made in understanding how cosmic structure forms from initial density 
fluctuations \cite{JPO}.
The IR data have revealed the `optically-hidden' side of galaxy 
formation, and shown that a significant fraction of the energy released by stellar 
nucleosynthesis is re-emitted as thermal radiation by dust. The underlying 
goal of all these efforts is to understand the growth of cosmic structures,
the internal properties of galaxies and their evolution, 
the mechanisms that shaped Hubble's morphological sequence, 
and ultimately to map   
the transition from the cosmic `dark age' to a ionized 
universe populated with luminous sources. While one of the important  
questions recently emerged is the nature (starbursts or active galactic 
nuclei?) and redshift distribution of the ultraluminous sub-mm sources 
discovered by {\it SCUBA}, of perhaps equal interest is the 
possible existence of a large population of faint galaxies still undetected 
at high redshifts, as the color-selected ground-based and 
{\it Hubble Deep Field} (HDF) samples include only the brightest and 
bluest star-forming objects. In any hierarchical clustering (`bottom-up')
scenario (the 
cold dark matter model being the best studied example), subgalactic
structures are the first non-linearities to form. High-$z$ 
dwarf galaxies and/or mini-quasars (i.e. an early generation of stars and 
accreting black holes in dark matter halos with circular velocities $v_c\sim 
50\,\kms$) may then be one of the main source of UV photons and 
heavy elements at early epochs. 

In this talk I will focus on some of the open issues and controversies 
surrounding our present understanding of the history of the conversion of 
cold gas into stars within galaxies, and of the evolution 
of luminous sources in the universe. An Einstein-deSitter (EdS) universe 
($\Omega_M=1$, $\Omega_\Lambda=0$) with $h=H_0/100\,\kmsmpc=0.5$ will be 
adopted in the following. 

\section*{Counting galaxies}
 
\begin{figure*}
\epsfysize=11cm 
\epsfxsize=10cm 
\hspace{3.5cm}\epsfbox{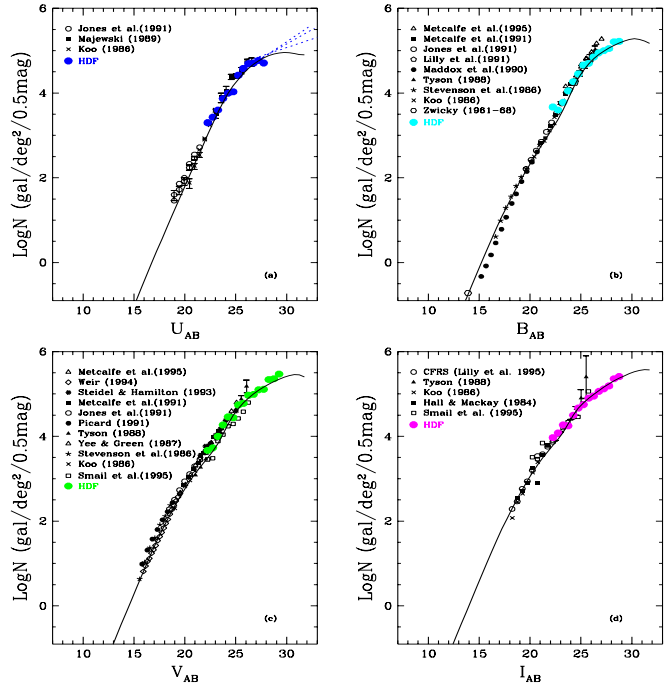}
\vspace{-0.8cm}
\caption[h]{\footnotesize Differential galaxy number counts per square degree 
as a function of apparent magnitude in four bandpasses from near-UV
to near-IR. The sources of the data points
are indicated in each panel (see \cite{Pozz98} for the complete
reference list). Note the decrease of the logarithmic slope $d\log
N/dm$ at faint magnitudes. 
}
\label{fig1}
\end{figure*}
Much observing time has been devoted in the past few years to the problem of
the detection of galaxies at high redshifts, as it was
anticipated that any knowledge of their early luminosity and color evolution
would set important constraints on the history of structure and star formation
in the universe.
As the best view to date of the optical sky at faint flux levels, the
HDF imaging survey has rapidly become a key testing ground
for models of galaxy evolution. The field, an undistinguished portion of the
northen sky at high galactic latitudes (the data from a southern deep field 
are being analyzed as we speak), is essentially a deep core 
sample of the universe, acquired with the {\it HST} in a 10-day exposure. With 
its depth -- reaching 5-$\sigma$ limiting AB
magnitudes of roughly 27.7, 28.6, 29.0, and 28.4 in $U, B, V,$ and $I$ 
\footnote{To get a feeling of the depth of this survey, note that 
$AB=29$ mag corresponds to the flux at Earth from a 
100 Watt light bulb at a distance of 10 million kilometers.}~ --
and four-filter strategy to provide constraints on the redshift and age 
distribution of galaxies in the image, the HDF
has offered the astronomical community the opportunity to study the galaxy 
population in unprecedented detail \cite{W96}.

There are about 3000 galaxies in the HDF, corresponding to $2\times 10^6$
deg$^{-2}$ down to the faint limit of the images. The galaxy counts 
are shown in Figure 1 in four bandpasses centered at 
roughly 300, 450, 600, and 800 nm.
A compilation of existing ground-based data is also shown, together with
the predictions of no-evolution models, i.e. models in which the absolute
brightness, volume density, and spectra of galaxies do not change with time. 
In all four bands, the logarithmic slope $\alpha$ of the galaxy number-apparent
magnitude counts, $\log N(m)=\alpha m$, flattens at faint magnitudes, e.g., from
$\alpha=0.45$ in the interval $21<B<25$ to $\alpha=0.17$ for $25<B<29$. 
The slope of the galaxy counts is a simple cosmological probe of the 
early history of star formation. The flattening at faint apparent magnitudes 
cannot be due to the reddening of distant sources as their Lyman break
gets redshifted into the blue passband,\footnote{For galaxies with $z>2$
($z>3.5$), the \HI Lyman edge shifts into the 300 (450) nm HDF 
bandpass. Neutral hydrogen, which is ubiquitous both within galaxies and in 
intergalactic space, strongly absorbs ultraviolet light, creating a 
spectral discontinuity that can be used to identify young, high-redshift
galaxies \cite{S96}.}~since the fraction of Lyman-break
galaxies at $B\sim 25$ is only of order 10\%.
Moreover, an absorption-induced loss of sources could not explain the similar
flattening of the galaxy counts observed in the $V$ and $I$ bands.
Rather, the change of slope suggests that the surface density of
luminous galaxies declines beyond $z\sim 1.5$. 

\begin{figure*}
\epsfysize=8cm 
\epsfxsize=8cm 
\hspace{3.5cm}\epsfbox{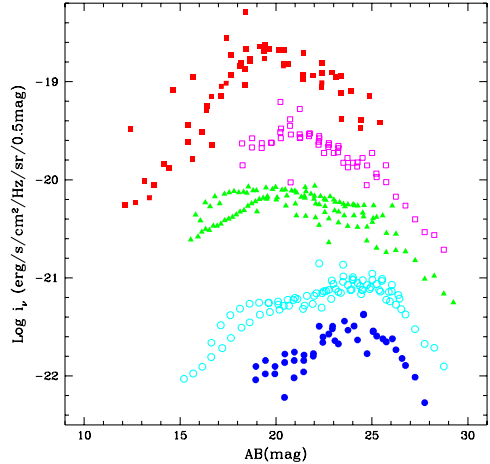}
\caption[h]{\footnotesize The contribution of known galaxies to the 
extragalactic background light per magnitude bin as a function of $U$ ({\it
filled circles}), $B$ ({\it open circles}), $V$ ({\it filled triangles}), $I$
({\it open squares}) and $K$ ({\it filled squares}) magnitudes. For clarity, 
the $B$, $V$, $I$ and $K$ values have been multiplied by a factor of 2, 8,
20, and 40, respectively.
}
\label{fig2}
\end{figure*}

\section*{The brightness of the night sky}

The extragalactic background light (EBL) is an indicator of the total 
luminosity of the universe. It provides unique information on the evolution 
of cosmic structures at all epochs, as the cumulative emission from galactic 
systems and active galactic nuclei (AGNs) is expected to be recorded in this 
background. 

The contribution of known galaxies to the optical EBL can be
calculated directly by integrating the emitted flux times the differential
galaxy number counts down to the detection threshold. The leveling off of the
counts is clearly seen in Figure 2, where the function
$i_\nu=10^{-0.4(m+48.6)}\times N(m)$ is plotted against apparent magnitude in
all bands \cite{Pozz98}. While counts having a logarithmic slope of
$\alpha\ge0.40$ continue to add to the EBL at the faintest magnitudes, it
appears that the HDF survey has achieved the sensitivity to capture the bulk of
the extragalactic light from discrete sources (an extrapolation of the observed
counts to brighter and/or fainter magnitudes would typically increase the
sky brightness by less than 20\%). To $AB=29$ mag, the sky brightness from
resolved galaxies in the $I$-band is $\approx 2\times 10^{-20}\uvunits$,
increasing roughly as $\lambda^2$ from 2000 to 8000 \AA.
The  flattening of the number counts has
the interesting consequence that the galaxies that produce $\sim 60\%$ of the
blue EBL have $B<24.5$. They are then bright enough to be identified in
spectroscopic surveys, and are indeed known to have median redshift $\langle
z\rangle=0.6$ \cite{Li96}. The quite general conclusion is that there 
is no evidence in the number-magnitude relation down to very faint flux 
levels for a large amount of star formation at high redshift. Note that
these considerations do not constrain the {\it rate} of starbirth at early
epochs, only the total (integrated over cosmic time) amount of stars -- hence
background light -- being produced, and neglect the effect of dust reddening. 

Figure 3 shows the total optical
EBL from known galaxies together with the recent {\it COBE} results.
The value derived by integrating the galaxy counts \cite{Pozz98} down to very
faint magnitude levels [because of the flattening at faint 
magnitudes of the number-magnitude relation most of the contribution to the 
optical EBL comes from relatively bright galaxies] 
implies a lower limit to the EBL intensity in the 
0.3--2.2 $\mu$m interval of $I_{\rm opt}\approx 12\,\eblunits$.\footnote{The 
direct detection of the optical EBL at 3000, 5500, and 8000 \AA\ 
derived from {\it HST} data \cite{RAB} implies values that are about a 
factor of two higher than the integrated light from galaxy counts.}~ When 
combined with the FIRAS and DIRBE measurements 
($I_{\rm FIR}\approx 16\,\eblunits$ in the 125--5000 $\mu$m range), this gives
an observed EBL intensity in excess of $28\,\eblunits$. 
The correction factor needed to account for the residual emission in the 2.2
to 125 $\mu$m region is probably $\lta2$ \cite{Dwe98}. (We shall see below 
how a population of dusty AGNs could make a significant 
contribution to the FIR background.) In the rest of this talk I will adopt a 
conservative reference value for the total EBL intensity associated 
with star formation activity over the entire history of the universe of 
$I_{\rm EBL}=40\,I_{40}\, \eblunits$.
\begin{figure*}
\epsfysize=7cm 
\epsfxsize=7cm 
\hspace{3.5cm}\epsfbox{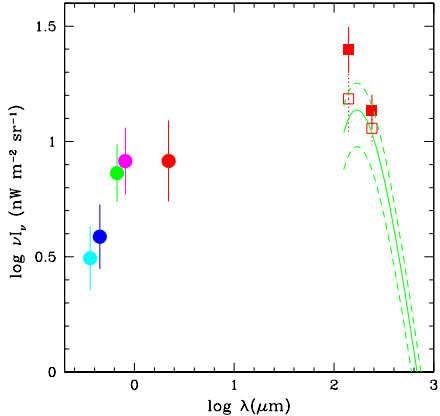}
\caption[h]{\footnotesize Spectrum of the extragalactic background light  
as derived from a compilation of ground-based and space-based galaxy counts in
the $U, B, V, I,$ and $K$-bands ({\it filled dots}), together with the FIRAS 
125--5000 $\mu$m ({\it solid and dashed lines}) and DIRBE 140 and 240 $\mu$m 
({\it filled squares}) detections \cite{Fix98}, \cite{Ha98}. The 
{\it empty squares} show the DIRBE
points after correction for WIM dust emission \cite{Lag}. 
\label{fig3}}
\end{figure*}

\section*{Modeling galaxy evolution}

In the past few years two different approaches have been widely used to
interpret faint galaxy data \cite{RSE}. In the
simplest version of what I will call the `traditional' scheme, 
a one-to-one mapping between
galaxies at the present epoch and their distant counterparts is assumed: one
starts from the local measurements of the distribution of galaxies as a
function of luminosity and Hubble type and models their photometric evolution
assuming some redshift of formation and a set of parameterized star formation
histories \cite{Tin80}. These, together with an initial mass function (IMF)
and a cosmological model, are then adjusted to match the observed number
counts, colors, and redshift distributions. Beyond the intrinsic simplicity of
assuming a well defined collapse epoch and pure-luminosity evolution
thereafter, the main advantage of this kind of approach is that it can easily
be made consistent with the classical view that ellipticals and spiral
galaxy bulges (both redder than spiral disks and containing less gas) formed 
early in a single burst of duration 1 Gyr or less 
\cite{Bern}. Spiral galaxies, by contrast, are characterized by a slower 
metabolism, i.e. star formation histories that extend to the present epoch.
In these models, typically, much of the action happens at 
high-redshifts.

A more physically motivated way to interpret the observations is to construct
semianalytic hierarchical models of galaxy formation and evolution 
\cite{WF91}. Here, one starts ab initio 
from a power spectrum of primordial density 
fluctuations, and follows the formation and hierarchical merging of the 
dark matter halos that provide the early seeds for later galaxy formation. 
Baryonic gas gets accreted onto the halos and is shock-heated.
Various prescriptions for gas cooling, star formation, feedback, and dynamical
friction are adopted, and tuned to match the statistical properties of 
both nearby
and distant galaxies.  In this scenario, there is no period when bulges and
ellipticals form rapidly as single units and are very bright: rather, small
objects form first and merge continually to make larger ones. Galaxy do not
evolve as isolated objects, and the rate of interaction was higher in the past. 
The bulk of the galaxy population is predicted to have been assembled quite 
recently, and most galaxies never experience star formation rates in excess
of a few solar masses per year.

\section*{Star formation history}

\begin{figure*}
\epsfysize=10cm 
\epsfxsize=10cm 
\hspace{3.5cm}\epsfbox{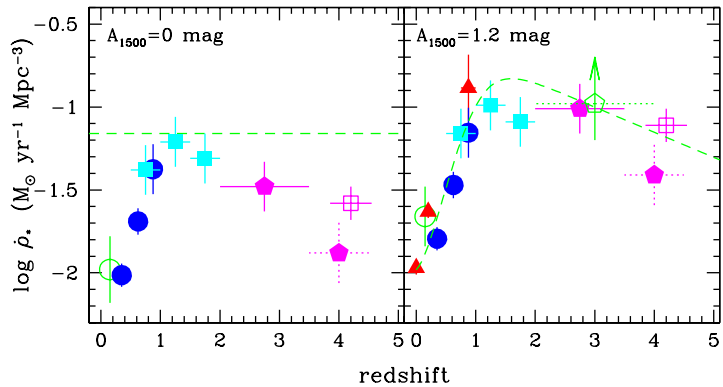}
\vspace{-4.cm}
\caption[h]{\footnotesize {\it Left}: Mean comoving density of star formation as 
a function of cosmic time. The data points with error bars have been inferred
from the UV-continuum luminosity densities of \cite{Li96} ({\it filled dots}),
\cite{Co97} ({\it filled squares}), \cite{M98} ({\it filled pentagons}), 
\cite{Treyer98} ({\it empty dot}), and \cite{Ste98} ({\it empty square}). 
The {\it dotted line} shows the fiducial rate, $\langle \dot{\rho_*}\rangle
=0.054\,\sfrd$, required to generate the observed EBL. {\it Right}: dust 
corrected values ($A_{1500}=1.2$ mag). The H$\alpha$ determinations of 
\cite{Ga95}, \cite{TM98}, and \cite{Gla98} ({\it filled triangles}), 
together with the {\it SCUBA} lower limit \cite{Hu98} ({\it empty pentagon}) 
have been added for comparison.
\label{fig4}}
\end{figure*}

Recently, it has become familiar to follow an alternative method, which 
focuses on the emission properties of the galaxy population as a whole. 
It traces the cosmic
evolution with redshift of the galaxy luminosity density and offers the
prospect of an empirical determination of the global star formation history of
the universe and IMF of stars independently of the merging histories, 
complex evolutionary phases, and possibly short-lived star formation
episodes of individual galaxies. The
technique relies on two basic properties of stellar populations: a) the
UV-continuum emission in all but the oldest galaxies is dominated by
short-lived massive stars, and is therefore a direct measure, for a given IMF
and dust content, of the instantaneous star formation rate; and b) the
rest-frame near-IR light is dominated by near-solar mass evolved stars, the
progenitors of which make up the bulk of a galaxy's stellar mass, and 
is more sensitive to the past star-formation history than the blue (and UV)
light. By modeling the ``emission history'' of the
universe at ultraviolet, optical, and near-infrared wavelengths from the
present epoch to high redshifts, one should be able to shed light on some 
key questions in
galaxy formation and evolution studies: Is there a characteristic epoch of star
and metal formation in galaxies?  What fraction of the luminous baryons
observed today were already locked into galaxies at early epochs? Are high-$z$
galaxies obscured by dust? Do spheroids form early and rapidly? Is there a
universal IMF?
 
The comoving volume-averaged history of star formation follows a  
relatively simple dependence on redshift. Its latest version, uncorrected for 
dust extinction, is plotted in Figure 4 ({\it left}). The measurements 
are based upon
the rest-frame UV luminosity function (at 1500 and 2800 \AA), assumed
to be from young stellar populations \cite{M96}. The prescription for a 
`correct'
de-reddening of these values has been the subject of an ongoing debate.   
Dust may play a role in obscuring the UV continuum of Canada-France 
Reshift Survey (CFRS, $0.3<z<1$) and Lyman-break ($z\approx 3$) galaxies, 
as their colors are too red to be fitted with an evolving stellar 
population and a Salpeter IMF \cite{M98}. Figure 
4 ({\it right}) depicts an extinction-corrected version of the same plot. 
The best-fit cosmic star formation 
history (shown by the dashed-line) with such a universal correction
produces a total EBL of $37\,
\eblunits$. About 65\% of this is radiated in the UV$+$optical$+$near-IR
between 0.1 and 5 $\mu$m; the total amount of starlight that is 
absorbed by dust and reprocessed in the far-IR is $13\,\eblunits$. 
Because of the uncertainties associated with the incompleteness of  
the data sets, photometric redshift technique, dust reddening, and UV-to-SFR 
conversion, these numbers are only meant to be indicative. On the other
hand, this very simple model is not in obvious disagreement with any of the observations, 
and is able, in particular, to provide a reasonable estimate of the 
galaxy optical and near-IR luminosity density. 

\section*{The stellar baryon budget}

With the help of some simple stellar population synthesis tools it is 
possible at this stage to make an estimate of the stellar mass density that
produced the integrated light observed today. The total {\it bolometric} 
luminosity of a simple stellar 
population (a single generation of coeval, chemically homogeneous
stars) having mass $M$ 
can be well approximated by a power-law with time for all ages $t\gta 100$ 
Myr, 
\begin{equation}
L(t)=1.3\,L_\odot {M\over M_\odot} \left({t\over 1\,{\rm Gyr}}\right)^{-0.8}
\end{equation}
(cf. \cite{Bu95}), 
where we have assumed solar metallicity and a Salpeter IMF truncated at 0.1
and 125 $M_\odot$. In a stellar system with arbitrary star formation rate per
unit cosmological volume, $\dot \rho_*$, the comoving bolometric emissivity 
at time $t$ is given by the convolution integral
\begin{equation}
\rho_{\rm bol}(t)=\int_0^t L(\tau)\dot \rho_*(t-\tau)d\tau.
\end{equation}
The total background light observed at Earth ($t=t_H$) is 
\begin{equation}
I_{\rm EBL}={c\over 4\pi} \int_0^{t_H} {\rho_{\rm bol}(t)\over 1+z}dt,
\end{equation}
where the factor $(1+z)$ at the denominator is lost to cosmic expansion 
when converting from observed to radiated (comoving) luminosity density. 
From the above equations it is easy to derive in a EdS cosmology 
\begin{equation}
I_{\rm EBL}=740\,\eblunits \langle {\dot\rho_*\over \sfrd}\rangle 
\left({t_H\over 13\,{\rm Gyr}}\right)^{1.87}.
\end{equation}
The observations shown in Figure 3 therefore imply a ``fiducial'' mean star 
formation density of $\langle \dot\rho_*\rangle=0.054\, I_{40}\,\sfrd$. 
The total stellar mass density observed today is 
\begin{equation}
\rho_*(t_H)=(1-R)\int_0^{t_H} \dot \rho_*(t)dt\approx 5\times 10^8\,I_{40}\
\mden 
\end{equation}
(corresponding to $\Omega_*=0.007\,I_{40}$), where $R$ is the mass fraction 
of a generation of stars that is returned to the interstellar medium, 
$R\approx 0.3$ for a Salpeter IMF. The optical/FIR background 
therefore requires that about 10\% of the nucleosynthetic baryons today 
\cite{Bur98} are in the forms of stars and their remnants.
The predicted stellar mass-to-blue light ratio is $\langle 
M/L_B\rangle\approx 5$. These values are quite sensitive to the
lower-mass cutoff of the IMF, as very-low mass stars can contribute
significantly to the mass but not to the integrated light of the whole stellar
population. A lower cutoff of 0.5$\msun$ instead of the 0.1$\msun$ adopted
would decrease the mass-to-light ratio (and $\Omega_*$) by a factor of 1.9 
for a Salpeter function. 

\section*{Two extreme scenarios}

\begin{figure*}
\epsfysize=10cm 
\epsfxsize=10cm 
\hspace{3.5cm}\epsfbox{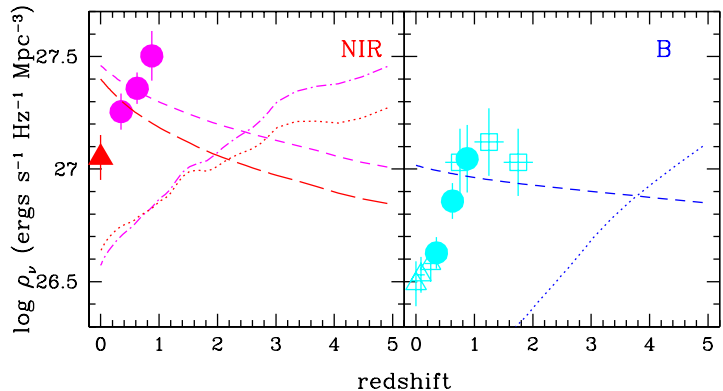}
\vspace{-4.0cm}
\caption[h]{\footnotesize {\it Left}: Synthetic evolution  of the near-IR 
luminosity density at rest-frame wavelengths of 1.0 ({\it long-dashed line}) and 2.2 
$\mu$m ({\it short-dashed line}). The model assumes a constant star formation rate of 
$\dot{\rho_*}=0.054\,\sfrd$ (Salpeter IMF). The {\it dotted} (2.2 $\mu$m) and {\it 
dash-dotted} (1.0 $\mu$m) curves show the emissivity of a simple stellar 
population with formation redshift $z_{\rm on}=5$, and total mass equal to
the mass observed in spheroids today \cite{Fuk98}.    
The data points are taken from \cite{Li96} 
({\it filled dots}) and \cite{Ga97} ({\it filled triangle}). 
{\it Right}: Same but in the $B$-band. The data points are taken from \cite{Li96} 
({\it filled dots}),  \cite{El96} ({\it empty triangles}), and \cite{Co97} 
({\it empty squares}). 
\label{fig5}}
\end{figure*}

Based on the agreement between the $z\approx 3$ and $z\approx 4$ luminosity 
functions at the bright end, it has been recently argued \cite{Ste98} that 
the decline in the luminosity density of faint HDF Lyman-break galaxies 
observed in the same redshift interval \cite{M96} may not be real, but simply 
due to sample variance in the HDF. When extinction corrections are applied, 
the emissivity per unit comoving volume due to star formation may then 
remain essentially flat for all redshift $z\gta 1$ (see Fig. 4). While this
has obvious implications for hierarchical models of structure formation,
the epoch of first light, and the reionization of the intergalactic medium 
(IGM), it is also interesting to speculate on the possibility of a constant
star-formation density at {\it all} epochs $0\le z\le 5$. 
Figure 5 shows the time evolution of the blue and near-IR
rest-frame luminosity density of a stellar population characterized by a 
Salpeter IMF, solar metallicity, and a (constant) star formation rate of 
$\dot{\rho_*}=0.054\,\sfrd$ (needed to produce the observed EBL).
The predicted evolution 
appears to be a poor match to the observations: it overpredicts the 
local $B$ and $K$-band luminosity densities, and underpredicts the 1$\,\mu$m 
emissivity at $z\approx 1$. 

At the other extreme, we know from stellar population studies that about half
of the present-day stars are contained in spheroidal systems, i.e. elliptical 
galaxies and spiral galaxy bulges, and that these stars formed early and 
rapidly. The expected rest-frame blue and near-IR emissivity
of a simple stellar population with formation redshift $z_{\rm on}=5$ and 
total mass density equal to the mass in spheroids observed today 
is shown in Figure 5. In this model the near-IR and blue
emissivities at $z=4-5$ are comparable with the values observed at $z=1$.
{\it HST}-NICMOS deep observations may be 
able to test similar scenarios for the formation of elliptical galaxies at 
early times.  

\section*{The mass density in black holes}

Recent dynamical evidence indicates that supermassive black holes reside 
at the center of most nearby galaxies. The available data (about 30 objects) 
show a strong correlation (but with a large scatter) between bulge and black 
hole mass \cite{Mag98}, with $M_{\rm bh}=0.006 \, M_{\rm bulge}$ as a 
best-fit. The total mass density in spheroids today is $\Omega_{\rm bulge}=
0.0036^{+0.0024}_{-0.0017}$ \cite{Fuk98}, implying a mean mass density of
dead quasars 
\begin{equation}
\rho_{\rm bh}=1.5^{+1.0}_{-0.7}\times 10^6\,\mden.
\end{equation}
Since the observed energy density from all quasars is equal to the
emitted energy divided by the average quasar redshift \cite{Sol82}, the 
total contribution to the EBL from accretion onto black holes is 
\begin{equation}
I_{\rm bh}={c^3\over 4\pi} {\eta \rho_{\rm bh}\over \langle 1+z\rangle}
\approx 18\,\eblunits \eta_{0.1}\langle 1+z\rangle^{-1},
\end{equation}
where $\eta_{0.1}$ is the efficiency for transforming accreted rest-mass 
energy into radiation (in units of 10\%). Quasars at 
$z\lta 2$ could then make a significant contribution to the brightness of the
night sky
if dust-obscured accretion onto supermassive black holes is an efficient
process \cite{Hae98}, \cite{Fab}. \footnote{It might be interesting to note 
in this context that a population of AGNs with strong intrinsic 
absorption (Type 
II quasars) is actually invoked in many current models for the X-ray 
background \cite{Mad94}, \cite{Com95}.} 

\section*{The end of the `dark ages'}

The epoch of reionization marked the end of the `dark ages' during which
the ever-fading primordial background radiation cooled below 3000 K and shifted
first into the infrared and then into the radio. Darkness persisted until
early structures collapsed and cooled, forming the first stars and quasars
that lit the universe up again \cite{Rees98}.
The application of the Gunn-Peterson constraint on 
the amount of smoothly distributed neutral material along the line of sight 
to distant objects requires the hydrogen component of the diffuse IGM to 
have been highly ionized by $z\approx 5$ \cite{SSG}, and the helium 
component by $z\approx 2.5$ \cite{DKZ}. From QSO absorption studies we also 
know that neutral hydrogen at high-$z$ accounts for only a small 
fraction, $\sim 10\%$, of the nucleosynthetic baryons \cite{LWT}. 

A substantial population of dwarf galaxies having star formation rates 
$<0.3\sfr$, and a space density in excess of
that predicted by extrapolating to faint magnitudes the best-fit
Schechter function, may be expected to form at early times in hierarchical 
clustering models, and has been recently proposed \cite{MR98}, \cite{MHR} 
as a possible candidate for photoionizing the IGM at these early
epochs. Establishing the character of
cosmological ionizing sources is an efficient way to constrain competing
models for structure formation in the universe, and to study the collapse and
cooling of small mass objects at early epochs. 

\begin{figure*}
\epsfysize=10cm 
\epsfxsize=10cm 
\hspace{3.5cm}\epsfbox{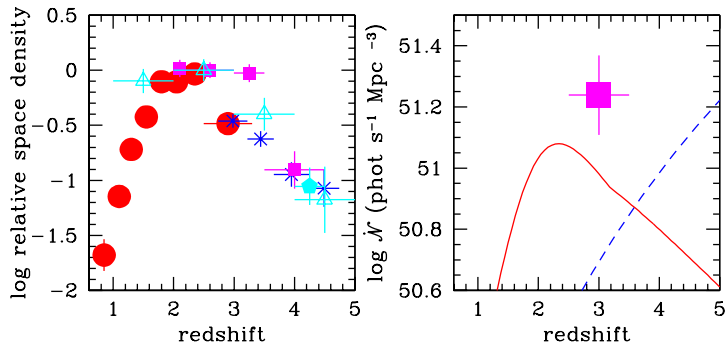}
\vspace{-4.0cm}
\caption[h]{\footnotesize {\it Left}: comoving space density of 
bright QSOs as a 
function of redshift. The data points with error bars are taken from 
\cite{HS90} {\it (filled dots)}, \cite{WHO} {\it (filled squares)}, 
\cite{Sc95} {\it (crosses)}, and \cite{KDC} {\it (filled pentagon)}. 
The {\it empty triangles} show the space density of radio-loud quasars 
\cite{H98}. {\it Right}: comoving emission rate of hydrogen Lyman-continuum 
photons ({\it solid line}) from QSOs, compared with the minimum rate 
({\it dashed line}) which is needed to fully ionize a fast recombining (with 
gas clumping factor $C=30$) EdS universe with 
$\Omega_bh^2=0.02$. Models based on photoionization by quasar sources 
appear to fall short at $z\approx 5$. 
The data point shows the estimated contribution from star-forming 
galaxies at $z\approx 3$, assuming that the fraction of Lyman continuum 
photons which escapes the galaxy \HI layers into the intergalactic medium 
is 50\% (see \cite{MHR} for details).
\label{fig6}}
\end{figure*}
The study of the candidate sources of ionization at $z=5$ can be simplified by 
noting that the {\it breakthrough epoch} (when all radiation sources can see
each other in the hydrogen Lyman-continuum) occurs much later in the 
universe than
the {\it overlap epoch} (when individual ionized zones become simply
connected and every point in space is exposed to ionizing radiation). 
This implies that at high redshifts the ionization equilibrium is actually 
determined by the {\it instantaneous } UV production rate \cite{MHR}. The 
fact that the IGM is rather clumpy and still optically thick at overlapping, 
coupled to recent
observations of a rapid decline in the space density of radio-loud quasars
and of a large population of star-forming galaxies at $z\gta 3$, has some 
interesting implications for rival ionization scenarios and for the star 
formation activity in the interval $<3<z<5$. 

The existence of a decline in the space density of bright quasars at redshifts
beyond $\sim 3$ was first suggested by \cite{O82}, and has been since then
the subject of a long-standing debate. In recent years, several optical 
surveys have consistently provided new evidence for a turnover in the QSO 
counts \cite{HS90}, \cite{WHO}, \cite{Sc95}, \cite{KDC}. 
The interpretation of the drop-off observed in optically selected samples is
equivocal, however, because of the possible bias introduced by dust 
obscuration arising from intervening systems. Radio emission, on the other 
hand, is unaffected by dust, and it has recently been shown \cite{Sha} that 
the space density of radio-loud quasars also decreases strongly for $z>3$. 
This argues that the turnover is indeed real and that dust along the line of 
sight has a minimal effect on optically-selected QSOs. In this case the 
QSO emission rate of hydrogen ionizing photons per unit comoving volume 
drops by a factor of 3 from $z=2.5$ to $z=5$, as shown in Figure 6.

Galaxies with ongoing star-formation are another obvious source of 
Lyman-continuum photons. Since the 
rest-frame UV continuum at 1500 \AA\ (redshifted into the visible band for a
source at $z\approx 3$) is dominated by the same short-lived, massive stars
which are responsible for the emission of photons shortward of the Lyman edge,
the needed conversion factor, about one ionizing photon every 10 photons at
1500 \AA, is fairly insensitive to the assumed IMF and independent of the
galaxy history for $t\gg 10^7\,$ yr. Figure 6 ({\it right}) shows the 
estimated Lyman-continuum luminosity density of galaxies at $z\approx 3$.
The data point assumes a value of 
$f_{\rm esc}=0.5$ for the unknown fraction of ionizing photons which escapes 
the galaxy \HI layers into the intergalactic medium. One 
should note that, while highly reddened galaxies at high 
redshifts would be missed by the Lyman-break color technique (which isolates 
sources that have blue colors in the optical and a sharp drop in the 
rest-frame UV), it seems unlikely that very dusty objects (with $f_{\rm esc}
\ll 1$) would contribute in any significant manner to the ionizing 
metagalactic flux. 

\section*{Reionization}

When an isolated point source of ionizing radiation turns on in a neutral
medium, the ionized
volume initially grows in size at a rate fixed by the emission of UV photons,
and an ionization front separating the \HII and \HI regions propagates
into the neutral gas. Most photons travel freely in the ionized bubble, and are
absorbed in a transition layer. The evolution of an expanding \HII region is 
governed by the equation 
\begin{equation}
{dV_I\over dt}-3HV_I={\dot N_{\rm ion}\over \bar{n}_\nH}-{V_I\over 
\bar{t}_{\rm rec}}, \label{eq:dVdt} 
\end{equation}
where $V_I$ is the proper
volume of the ionized zone, $\dot N_{\rm ion}$ is the number of ionizing 
photons emitted by the central source per unit time, $\bar{n}_\nH$ is the 
mean hydrogen density of the expanding IGM, $H$ is the Hubble constant, and 
$\bar{t}_{\rm rec}$ is the hydrogen mean recombination 
timescale, 
\begin{equation}
\bar{t}_{\rm rec}=[(1+2\chi) \bar{n}_\nH \alpha_B\,C]^{-1}=0.3\, {\rm Gyr}
\left({\Omega_b h^2 \over 0.02}\right)^{-1}\left({1+z\over 4}\right)^{-3}
C_{30}^{-1}. \label{eq:trec}
\end{equation}
One should point out that the use of a volume-averaged clumping factor, 
$C\equiv \langle n_\nHII^2\rangle/\bar{n}_\nHII^2$, in the 
recombination timescale 
is only justified when the size of the \HII region is large compared to the 
scale of the clumping, so that the effect of many clumps (filaments) within 
the ionized volume can be averaged over. The validity of this approximation
can be tested by numerical simulations (see Figure 7). 
Across the I-front the degree of
ionization changes sharply on a distance of the order of the mean free path of
an ionizing photon. When $\bar{t}_{\rm rec}$ is much smaller than the Hubble
time, the growth of the \HII region is slowed down by recombinations in 
the highly inhomogeneous medium, and its evolution
can be decoupled from the expansion of the universe. 

\begin{figure*}
\epsfysize=7cm 
\epsfxsize=7cm 
\hspace{3.5cm}\epsfbox{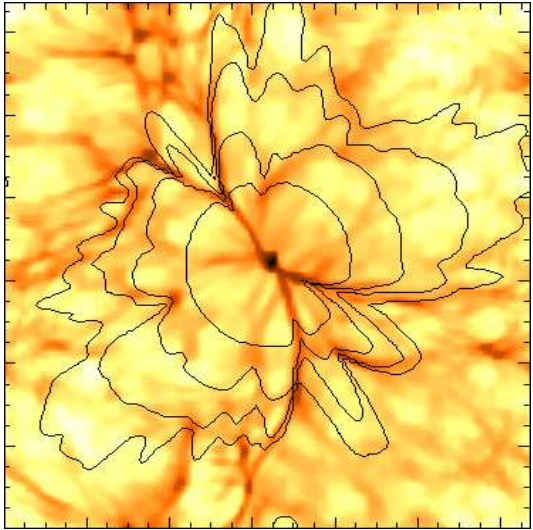}
\vspace{0.5cm}
\caption[h]{\footnotesize Simulating the reionization of the universe:
propagation of an ionization front in a $128^3$ 
cosmological density field. A `mini-quasar' with $\dot {N}=5\times 
10^{53}\,$ s$^{-1}$ was turned on at the densest cell, in a virialized halo 
of total mass $1.3\times 10^{11}\,M_\odot$. The box length is 2.4 comoving 
Mpc. The solid contours give the position of the front at 0.15, 0.25, 0.38, and
0.57\,Myr after the quasar has switched on at $z=7$. The underlying greyscale
image indicates the initial \HI density field. (From \cite{Ab98}.)
\label{fig7}}
\end{figure*}

In analogy with the individual \HII region case, it can be shown that the
hydrogen
component in a highly inhomogeneous universe is completely reionized when the
number of photons emitted above 1 ryd in one recombination time equals the
mean number of hydrogen atoms.  At any given epoch there is a critical value 
for the emission rate of ionizing photons per unit cosmological comoving 
volume,   
\begin{equation}
\dot {\cal N}_{\rm ion}(z)={\bar{n}_\nH(0)\over \bar{t}_{\rm rec}(z)}=(10^{51.2}\,
\ndotunits)\, C_{30} \left({1+z\over 6}\right)^{3}\left({\Omega_b 
h^2\over 0.02}\right)^2, 
\label{eq:caln}
\end{equation}
which is independent of the (unknown) previous emission history of 
the universe: only
rates above  this value will provide enough UV photons to ionize the IGM by 
that epoch. One can then compare our estimate of  $\dot {\cal N}_{\rm
ion}$ to the inferred contribution from QSOs and star-forming galaxies. 
The uncertainty on this critical rate is difficult to estimate, as it depends 
on the clumpiness of the IGM  (scaled in the expression above 
to the value inferred at $z=5$ from numerical simulations \cite{GO97})
and the nucleosynthesis constrained baryon density. The 
evolution of the critical rate as a function of redshift is plotted in Figure 
6 ({\it right}). While $\dot {\cal N}_{\rm ion}$ is comparable to the quasar 
contribution at 
$z\gta 3$, there is some indication of a deficit of Lyman-continuum 
photons at $z=5$. For bright, massive galaxies to produce enough UV 
radiation at
$z=5$, their space density would have to be comparable to the one observed at
$z\approx 3$, with most ionizing photons being able to escape freely from the
regions of star formation into the IGM. This scenario may be in 
conflict with  direct observations of local starbursts below
the Lyman limit showing that at most a few percent of the stellar ionizing
radiation produced by these luminous sources actually escapes into the IGM 
\cite{Le95}.\footnote{At $z=3$ Lyman-break galaxies radiate into the IGM 
more ionizing photons than QSOs if $f_{\rm esc}\gta 30\%$.}~

It is interesting to convert the derived value of $\dot {\cal N}_{\rm ion}$  
into a ``minimum'' star formation rate per unit (comoving) volume, $\dot 
\rho_*$:
\begin{equation}
{\dot \rho_*}(z)=\dot {\cal N}_{\rm ion}(z) \times 10^{-53.1} f_{\rm esc}^{-1}
\approx 0.013 f_{\rm esc}^{-1} \left({1+z\over 6}\right)^3\ \sfrd. \label{eq:sfr} 
\end{equation}  
The star formation density given in the equation above is comparable 
with the value directly ``observed''
(i.e., uncorrected for dust reddening) at $z\approx 3$ \cite{M98}.
The conversion factor assumes a Salpeter IMF with solar metallicity, and has
been computed using a population synthesis code \cite{BC98}.
It can be understood by noting that, for each 1 $M_\odot$ of stars formed,
8\% goes into massive stars with $M>20 M_\odot$ that dominate the
Lyman-continuum luminosity of a stellar population. At the end of the C-burning
phase, roughly half of the initial mass is converted into helium and carbon,
with a mass fraction released as radiation of 0.007. About 25\% of the energy
radiated away goes
into ionizing photons of mean energy 20 eV. For each 1 $M_\odot$ of stars
formed every year, we then expect
\begin{equation}
{0.08\times 0.5 \times 0.007 \times 0.25\times M_\odot c^2\over
20 {\,\rm eV\,}} {1\over  {\rm 1\, yr}} \sim 10^{53}\ndotun
\end{equation}
to be emitted shortward of 1 ryd. 

\section*{Conclusions}

Recent studies of the volume-averaged history of stellar birth 
are pointing to an era of intense star formation at $z\approx 1-1.5$.
The optical datasets imply that a fraction close to 65\% of the present-day 
stars was produced at $z>1$, and only 25\% at $z>2$. About
half of the stars observed today would be more than 9 Gyr old, and only 10\%
would be younger than 5 Gyr.\footnote{Unlike the measured
number densities of objects and rates of star formation, the integrated stellar
mass density does not depend on the assumed cosmological model.}~
There is no `single epoch of galaxy formation': rather, it appears that
galaxy formation is a gradual process.
Numerous uncertainties remain, however, particularly the role played
by dust in obscuring star-forming objects. 
Our first glimpse of the history of galaxies
to $z\sim 4$ leads to the exciting question of what happened before.	
Substantial sources of ultraviolet photons must 
have been present at $z\gta 5$ to keep the universe ionized, 
perhaps low-luminosity quasars \cite{HL98} or a first generation of stars in 
dark matter halos with virial temperature $T_{\rm vir}\sim 10^4-10^{5} \,$K 
\cite{OG96}, \cite{HL97}. Early star
formation provides a possible explanation for the widespread existence of heavy
elements in the \Lya forest \cite{Cow95}, while reionization by QSOs may
produce a detectable signal in the radio extragalactic background at meter
wavelengths \cite{Mad97}. A detailed exploration of such territories must
await projected facilities like the {\it Next Generation Space Telescope}  
and the {\it Square Kilometer Radio Telescope}.
 
\bigskip\bigskip

{\bf Acknowledgments}
\bigskip

Support for this work was provided by NASA through ATP grant NAG5--4236.

\vfill
\end{document}